# PANDA: A self-driving lab for studying electrodeposited polymer films

Harley Quinn,[a] Gregory A. Robben,[a] Zhaoyi Zheng,[a] Alan L. Gardner,[b] Jörg G. Werner,*[abc] and Keith A. Brown*[abd]

[a]Division of Materials Science & Engineering, Boston University, Boston, MA, 02215, USA.

[b]Department of Mechanical Engineering, Boston University, Boston, MA, 02215, USA

[c]Department of Chemistry, Boston University, Boston, MA, 02215, USA

[d]Department of Physics, Boston University, Boston, MA, 02215, USA

*E-mail: brownka@bu.edu, jgwerner@bu.edu

# Abstract

We introduce the polymer analysis and discovery array (PANDA), an automated system for high-throughput electrodeposition and functional characterization of polymer films. The PANDA is a custom, modular, and low-cost system based on a CNC gantry that we have modified to include a syringe pump, potentiostat, and camera with a telecentric lens. This system can perform fluid handling, electrochemistry, and transmission optical measurements on samples in custom 96-well plates that feature transparent and conducting bottoms. We begin by validating this platform through a series of control fluid handling and electrochemistry experiments to quantify the repeatability, lack of cross-contamination, and accuracy of the system. As a proof-of-concept experimental campaign to study the functional properties of a model polymer film, we optimize the electrochromic switching of electrodeposited poly(3,4-ethylenedioxythiophene):poly(styrene sulfonate) (PEDOT:PSS) films. In particular, we explore the monomer concentration, deposition time, and deposition voltage using an array of experiments selected by Latin hypercube sampling. Subsequently, we run an active learning campaign based upon Bayesian optimization to find the processing conditions that lead to the highest electrochromic switching of PEDOT:PSS. This self-driving lab integrates optical and electrochemical characterization to constitute a novel, automated approach for studying functional polymer films.

# Introduction

Self-driving labs (SDL) have emerged as an enabling tool to accelerate research in materials science.[1,2] These systems leverage robotics to perform high-throughput synthesis and characterization while using machine learning to select experiments to accelerate material discovery or optimization. In addition to their main advantage, SDLs can offer benefits over conventional research in terms of reproducibility, resources used per experiment, and collation of metadata. These systems have seen application in fields such as photovoltaics,[3] battery research,[4] semiconductor nanoparticles,[5] additive manufacturing,[6] catalysis,[7] and organic lasing materials.[8] As exemplified by the breadth of these applications, SDLs often take the form of bespoke systems to perform specific classes of experiments. There has been sustained effort to develop modular software tools that can apply to many different SDL formats, such as Coscientist[9] or ChemOS,[10] however, the application of SDLs in a given domain still requires innovations in the hardware specific to that space.

One area that has received a particularly large amount of development with regard to SDLs is electrochemistry. Early work leveraged the common microtiter plate format to realize systems that could measure the electrochemical response of any well by mounting a set of electrodes on a moving gantry.[11,12] This concept was further developed through the use of a communal working electrode and 96 parallel counter electrodes to allow for the whole plate to be characterized simultaneously.[13] Further modifications to the array format have allowed researchers to include the ability to perform photochemistry alongside electrochemistry.[14] Custom arrays have also been used with sustained electrochemical monitoring to study leeching for copper extraction.[15] An alternate approach to using arrays of wells is to use a single electrochemical reactor multiple times. This approach has been more conducive to realizing closed-loop formulation and testing with systems being developed to study electrocatalysts,[16] electrolytes,[17] or the mechanism of electrochemical reactions.[18] In parallel with these hardware advances, software tools specific to electrochemistry have emerged such as Hard Potato,[19] a Python library for automating electrochemical experiments, and ExpFlow, a graphical user interface for automated electrochemical experiments.[20]

Despite these innovations, one area of electrochemistry that has been largely unexplored is the electrodeposition of polymer films. Electrodeposition is a powerful approach for realizing functional ultrathin films,[21–23] however, the optimization has not yet taken advantage of the acceleration inherent to SDLs. While SDLs have been used to study polymer films prepared through other means such as spin coating,[24,25] spray-coating,[26] drop-casting,[27,28] printing,[27] and spontaneous solution spreading,[29] electrodeposition and electrochemical characterization have yet to receive substantial attention. Furthermore, many functional properties of electrodeposited films require multi-modal characterization, such as optical characterization, that is incompatible with previously studied electrochemical SDLs.

Here, we introduce an open-source SDL that electrodeposits and functionally characterizes polymer films using a combination of electrochemical and optical techniques (Figure 1). Experiments are performed in a novel well plate architecture in which the transparent bottom of each well constitutes the working electrode while enabling transmission optical characterization. We perform an extensive series of experiments to determine the precision and accuracy of fluid handling while ruling out the potential for cross-contamination. In addition, we validate the performance and reproducibility of the custom electrochemical cells. Finally, we benchmark the functional performance of this SDL by running a fully autonomous campaign to optimize the electrochromic performance of electrodeposited poly(3,4-ethylenedioxythiophene) (PEDOT).

Crucially, this system is low cost and open source, meaning that others can adopt or modify it to explore myriad properties for electrodeposited polymers.

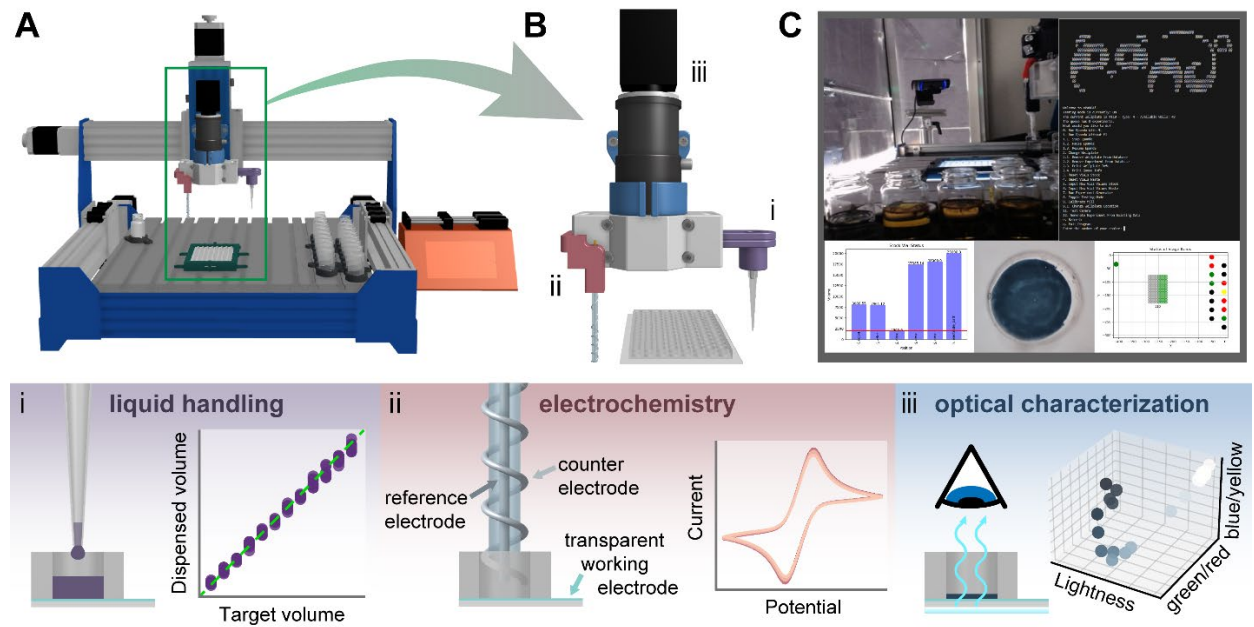

**Fig. 1** (A) A 3D rendering of the polymer analysis and discovery array (PANDA) highlighting the tool end and well plate in a green box. (B) Detailed 3D rendering of the tool end , featuring: (i) liquid handling attachment consisting of a custom pipette tip adapter, (ii) an electrochemical attachment enclosing a counter electrode and a reference electrode , and (iii) a telecentric lens attached to a camera for optical characterization. (C) Screenshot of PANDA interface showing real-time monitoring images (top left), main menu (top right), stock solution levels (bottom left), a captured image of an electrodeposited film (bottom center), and deck status (bottom right). (i) Schematic showing the liquid handling system and calibrated performance. (ii) Schematic showing the electrochemical system with example cyclic voltammetry data. (iii) Schematic showing optical characterization system with a depiction of how the studied films are perceived in CIE L*a*b* coordinates.

# Experimental

**Hardware**

The robotic system was built around a modified CNC router (Figure 1A - Genmitsu, PROVerXL 4030). The spindle was removed and replaced with a 3D printed modular tool holder that holds adapters for two electrodes of an electrochemical cell (Pt wire to serve as a counter electrode wrapped around a glass capillary housing an AgCl-coated Ag wire in a 1 M KCl solution capped with a glass frit to serve as a reference electrode), pipette tip adapter attached via a tube to a syringe pump, and telecentric lens (#52-271, Edmund optics) with attached c-mount camera (Grasshopper 3, FLIR) (Figure 1B). Accessories are attached to the deck of the mill using t-slots as mounting points to hold stock solutions, an electrolyte reference solution, waste vials, and a custom substrate mount (Figure S1). Mounted around the system are two process-monitoring cameras to allow visualization and recording of the key areas in the system including experiment progress and liquid handling. (Figure 1C) For liquid handling validation experiments, a piece of the deck was removed to allow the use of an analytical balance (Entris II Essential Precision Balance, Sartorius) while performing dispensing operations. For control of the fluidics, we integrated a syringe pump (Aladin Model A-1000, WPI) and for the electrochemistry a potentiostat (Interface 1010E, Gamry). Pictures detailing the system are included in Figure S2.

**Software**

The PANDA is controlled by a custom Python program that users interact with via a text-based terminal interface. Through this interface individual actions may be taken, such as updating the locations or contents of physical objects, generating experiment instruction sets, or initiating either a semiautonomous or fully autonomous campaign.

The program is a composition of custom scripts and modules which themselves depend upon the Python Standard Library and third-party open-source libraries such as gpytorch for machine learning. The overall architecture represents each virtual or physical system component in its own module in an effort to realize a modular and easy to maintain codebase. The individual modules are then used as configurable building blocks by higher level components responsible for orchestrating the mill's movements; experiment generation, selection, execution, and analysis; communication with the research team through Slack; control of Open Broadcaster Software for monitoring; and database communication. Further details of the software are included in the SI (Figures S3 – S6, Table S1) along with all code used in this work.

**Chemicals and materials**

For experiments to study fluid handling, de-ionized water (18.2 MΩ·cm Milli-Q, Millipore) was used. For experiments to study the electrochemical system, the PANDA used potassium ferricyanide (5 mM or 10 mM, 99+%, Thermo Scientific) in de-ionized water with potassium chloride (0.1 M, 99+%, Thermo Scientific) as the supporting electrolyte.

For the polymer film electrodeposition experiments, the deposition solution was made by dissolving 3,4-ethylenedioxythiophene (EDOT) (99%, Acros Organics) in a 1:1 (v:v) ratio of methanol (HPLC, Thermo Scientific) to de-ionized water at concentrations between 0.01 and 0.1 M and adding poly(styrene sulfonic acid) sodium salt (1 mM, M.W. 70,000 Da, Thermo Scientific). Lithium perchlorate (0.1 M, 99%, Sigma-Aldrich) was dissolved as the supporting electrolyte in de-ionized water and used for the oxidation and reduction of PEDOT films in the electrochromic characterization experiments.

**Working electrode fabrication and electrochemical measurements**

Glass slides (86 x 126 mm$^2$, polished borofloat glass – S.I. Howard Glass Co., Inc.) were cleaned in acetone followed by isopropyl alcohol (IPA) and dried under a nitrogen stream prior to use. A conductive layer of indium tin oxide (ITO) was sputtered onto the glass slides using DC Magnetron sputtering (Angstrom, EvoVac). The slides were subsequently annealed at 400 °C for 5 minutes resulting in a sheet resistivity of 410 Ω/sq.

Polydimethylsiloxane (PDMS) sheets were made by pouring a 10:1 base:crosslinker ratio of Sylgard 184 (Electron Microscopy Sciences) into a custom 3D printed polylactic acid (PLA) frame mounted on clean glass slides and then allowed to cure at room temperature for 48 hours. After removal from the glass substrate, wells were laser-cut (Epilog, Fusion Edge 12) into the PDMS sheets using a custom template designed to mimic the geometry of commercially available 96-well plates. After cutting, the PDMS was thoroughly cleaned with de-ionized water and then rinsed with IPA. The PDMS gasket was then mounted onto the ITO-coated glass by applying a thin layer of uncured PDMS to the gasket and aligning it onto the ITO-coated glass using an alignment tool.

All electrochemical experiments were conducted using a Gamry Interface 1010E potentiostat in a three-electrode configuration. The ITO-coated glass served as the working electrode, a platinum wire (diameter 0.25 mm, 99.9% trace metals basis) was used as the counter electrode, and an AgCl-coated Ag wire in a 1 M KCl solution separated from the working solution by a glass frit was used as the reference electrode (Ag/AgCl Reference Electrode, CH Instruments). The second cycle of cyclic voltammetry (CV) was used for all analyses to avoid any transient effects.

## Results and Discussion
### Validation of Automated Fluid Handling and Electrochemistry

In order to develop the confidence needed to utilize the PANDA as an SDL and perform polymer deposition experiments, it is first necessary to evaluate that each module of the system is functioning reliably. Foundational to any experiment is the ability to dispense fluid. As such, we began by exploring the accuracy and precision of PANDA-based fluid handling. Initially, we sought to compare the performance using each of the two conventional micro pipetting techniques: reverse pipetting and forward pipetting (Figure 2A). In reverse pipetting, extra fluid is drawn into the pipette with the intended volume specified by the dispensing step. In forward pipetting, the intended volume is determined by the amount drawn into the pipette with an additional "blow-out" volume of air being used to ensure that all fluid is dispensed. To explore the performance of each approach, we performed 600 randomized experiments in which 100 µL was pipetted onto an analytical balance by the PANDA (Figure 2B). Interestingly, we observed that forward pipetting produced more accurate results with the average dispensed quantity being 94.8 µL vs 70.6 µL for reverse pipetting. While inaccuracies may on average be corrected, perhaps more important was the difference in precision with forward pipetting producing a more precise standard deviation of 1.7 µL vs 4.0 µL for reverse pipetting. Based on these results, all subsequent work utilized forward pipetting.

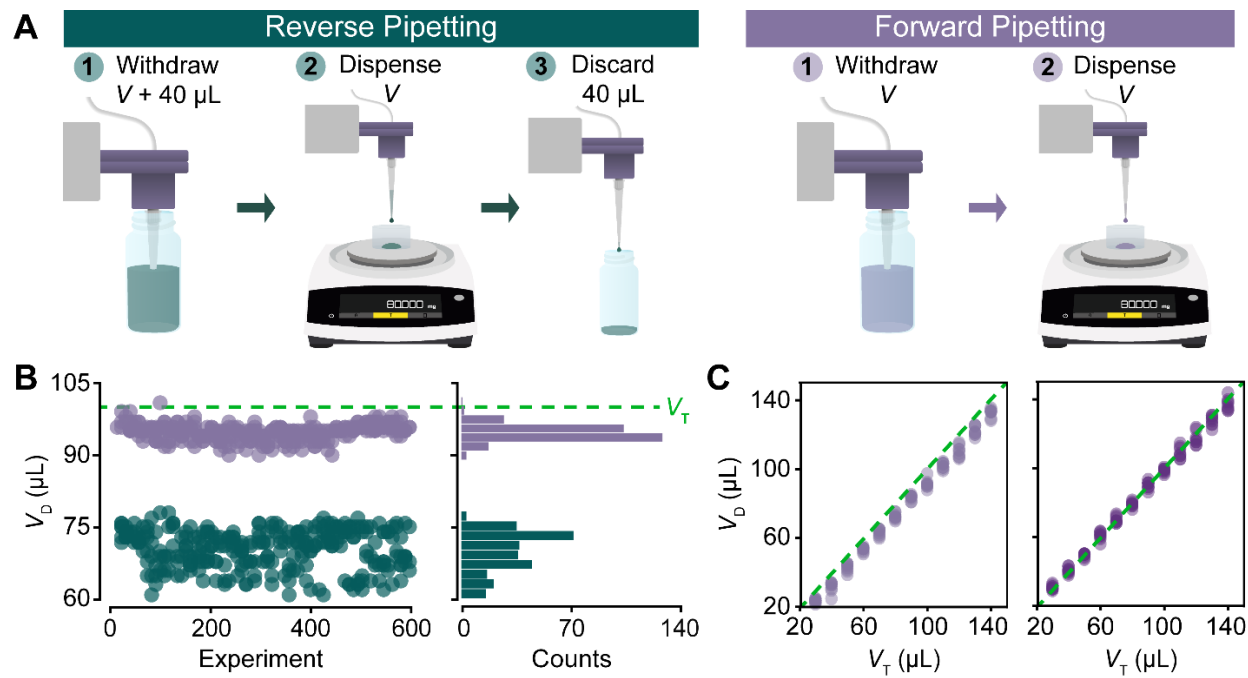

**Fig. 2** (A) Left schematic: Reverse pipetting process illustrated in three steps— (1) withdraw a programmed volume plus an additional 40 μL from the stock solution, (2) dispense the programmed volume into a container on an integrated analytical balance, (3) discard the remaining volume in the pipette tip. Right schematic: Forward pipetting process detailed in two steps— (1) withdraw a programmed volume from the stock solution, (2) dispense the entire solution into a container on an integrated analytical balance. (B) Comparison of volume dispensed $V_D$ when reverse pipetting (teal) and forward pipetting (purple) across 600 experiments. Histograms show the frequency of $V_D$ for each method. The 100 μL target volume $V_T$ is marked by a dashed green line. (C) Left: $V_D$ vs. $V_T$ over volumes ranging from 20 to 140 μL in 10 μL increments, with 8 replicates per volume. Right: Result of a second set of experiments after calibrating with a linear shift.

Having selected forward pipetting due to its comparatively high precision, we sought to determine a calibration strategy to improve the accuracy of fluid handling. In a calibration experiment, we dispensed volumes between 30 and 140 μL in 10 μL increments across eight replicates per volume while measuring the dispensed mass using the scale (Figure 2C left). The root mean squared error (RMSE) for the parity line (green) was calculated to be 7.8 μL, providing a measure of the accuracy without calibration. Fitting the data to a line, we find an RMSE of 2.3 μL, consistent with the precision reported in the prior study. Based on the linear appearance of the data, we hypothesized that applying a linear calibration function to adjust the programmed volume $V_T$ would remove any systematic inaccuracies. Thus, we repeated the experiments with the newly calibrated volumes (Figure 2C right). In the calibrated experiments, our accuracy improved to 2.8 µL, and our precision remained consistent at 2.3 µL, demonstrating a clear improvement in accuracy while maintaining precision with the application of a linear calibration function. To ensure accuracy throughout experiments we programmed the PANDA to perform 1000 withdrawals and dispensing actions with three different pipette tips and found minimal tip-to-tip variation (Figure S7). This allowed us to replace the pipette tip less frequently, saving time and minimizing waste.

Having calibrated the liquid handling system, we validated the electrochemical system next. While the materials that constitute the electrodes are relatively common, the geometry of the fluid cell is comparatively novel, meaning that well-to-well variability and repeatability should be assessed. Thus, we performed a series of experiments using aqueous $K_3Fe(CN)_6$ (FC), a standard redox probe,[30] in a series of experiments with KCl as the supporting electrolyte. We programmed the PANDA to deposit 120 µL of the FC solution into a single well, perform CV (operating voltage between -0.2 and 0.6 V vs. Ag/AgCl, scan rate 50 mV s$^{-1}$), remove the used solution, and then replicate this process for a total of ten experiments (Figure 3A). The CV results (Figure 3B) showed consistent anodic peak currents $j_{pa}$, peak current differences $\Delta j_p$, and peak separations $\Delta E_p$, indicating that the system is reproducible. This procedure was repeated in two additional wells to test well-to-well variability, results of which are shown in Figure 3C. Importantly, the mean results for the three wells were within 6 mV and 0.03 mA cm$^{-2}$ for their $\Delta E_p$ and $\Delta j_p$ values, respectively.

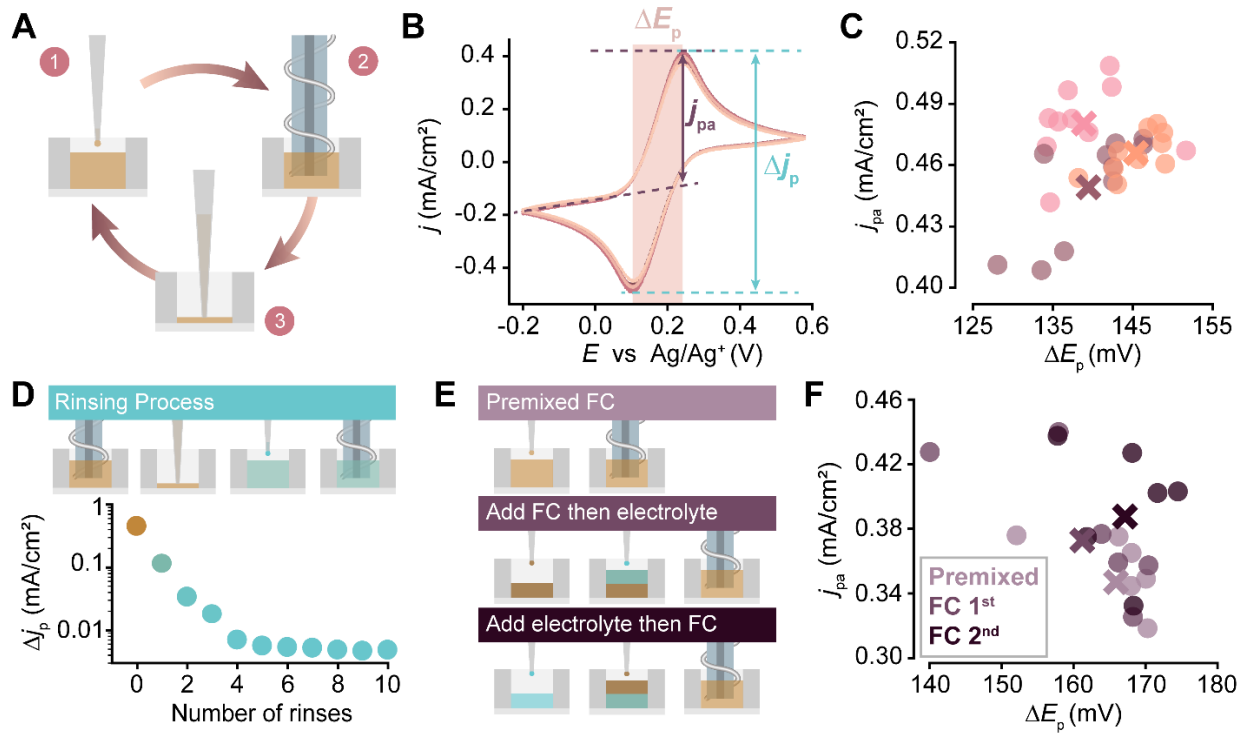

**Fig. 3** (A) Schematic of the K$_3$Fe(CN)$_6$ (FC) experiment process: (1) FC solution is added to a well, (2) cyclic voltammetry (CV) is performed, and (3) the solution is removed. (B) Ten CVs showing current density $j$ vs. potential $E$ vs. an Ag/Ag$^+$ reference electrode. Key features are highlighted: peak-to-peak separation $\Delta E_p$ in a shaded peach region, anodic peak height $j_{pa}$ between two dashed purple lines, and distance between peaks $\Delta j_p$ between two dashed blue lines. (C) Plot of $j_{pa}$ vs. $\Delta E_p$ from B with data from two additional wells with 10 experiments each. Marker color indicates the well used, with the X symbols denoting the averages for each well. (D) Schematic of the process used to rinse between experiments in the same well: (1) CV with FC, (2) removal of FC solution, (3) addition of electrolyte rinse, and (4) CV in the rinse solution. The result is shown as the bottom panel as $\Delta j_p$ vs. number of rinses with points colored by the residual FC concentration. (E) Schematics of the mixing strategies used in experiments: Top—5 mM FC dispensed directly; Middle—10 mM FC added and then diluted to 5 mM with electrolyte; Bottom—Electrolyte added first and then concentrated to 5 mM FC by adding 10 mM FC. In all experiments, a CV was performed after fluid handling. (F) Plot of $j_{pa}$ vs. $\Delta E_p$ resulting from the mixing strategies in E. Marker color indicates the mixing sequence as shown in E. The X symbols show the averages found for each method.

As many workflows of polymer analysis will involve multiple fluid handling steps, it is necessary to evaluate the degree to which both the pipette tip and wells can be rinsed without residual solution remaining. The capability of performing sequential steps in a well is particularly important for in-loop film characterization without human intervention. Thus, we programmed the PANDA to withdraw FC solution, dispense it directly into waste, and then perform a rinsing action (withdrawing and dispensing) three times before withdrawing electrolyte solution from a different stock vial, dispensing it into a new well, and performing CV to determine the presence of any residual redox-active species (Figure S8). After three rinses, residual redox-active species were undetectable, establishing the pipette rinse protocol for subsequent experiments. For the evaluation of our well rinsing procedure, the PANDA was programmed to dispense FC solution into a well, execute a CV, remove the solution from the well, fill the well with 120 µL of electrolyte solution, and then perform CV again to determine the presence of any residual redox-active species over multiple repetitions (Figure 3D top). Importantly, redox-active species were undetectable after four rinses, establishing the rinse protocol for subsequent experiments (Figure 3D bottom). Based on the fitting of the linear portion of this curve, approximately 75% of the redox-active species is removed from the well with each rinsing cycle.

As a final set of experiments to validate the performance of the PANDA, we sought to evaluate its ability to mix solutions in the wells. This capability is critical for studying series of reagent concentrations using a combination of finite stock solutions. To explore this, we compared the CVs that resulted from three different processes that should have resulted in the same final concentration: (1) a premixed control that was 5 mM FC, (2) wells prepared by first depositing 10 mM FC and then an equal volume of electrolyte, and (3) wells prepared by first depositing an electrolyte solution and then 10 mM FC (Figure 3E). A total of 18 CVs were performed in two separate wells with each well being rinsed after each experiment. Aside from the natural agitation arising from deposition, no further mixing was performed. The outcomes, depicted as $j_{pa}$ versus $\Delta E_p$, indicated that all preparation methods produced outcomes that were not statistically distinguishable (single factor ANOVA, α=0.05, p-value = 0.45 for $\Delta E_p$ and 0.14 for $j_{pa}$), confirming that the solutions were adequately mixed (Figure 3F).

**Building PEDOT training data set**

Following the successful validation of our liquid handling and electrochemical systems, we programmed the PANDA to electrodeposit PEDOT:PSS films as a proof of concept. Each experiment was a nine-step process (Figure 4A) in which the PANDA, (1) selected a previously unused well, (2) dispensed an EDOT solution into that well, (3) conducted electrodeposition using potentiostatic chronoamperometry, (4) rinsed the well with electrolyte solution, (5) imaged the deposited film, (6) dispensed the inert supporting electrolyte solution for switching the electrochromic state, (7) reduced the film by applying a negative potential, (8) rinsed the well with electrolyte solution, and then (9) imaged the film in its bleached state (Figure 4A). A Flowchart detailing the software steps used by the PANDA are shown in Figure S9.

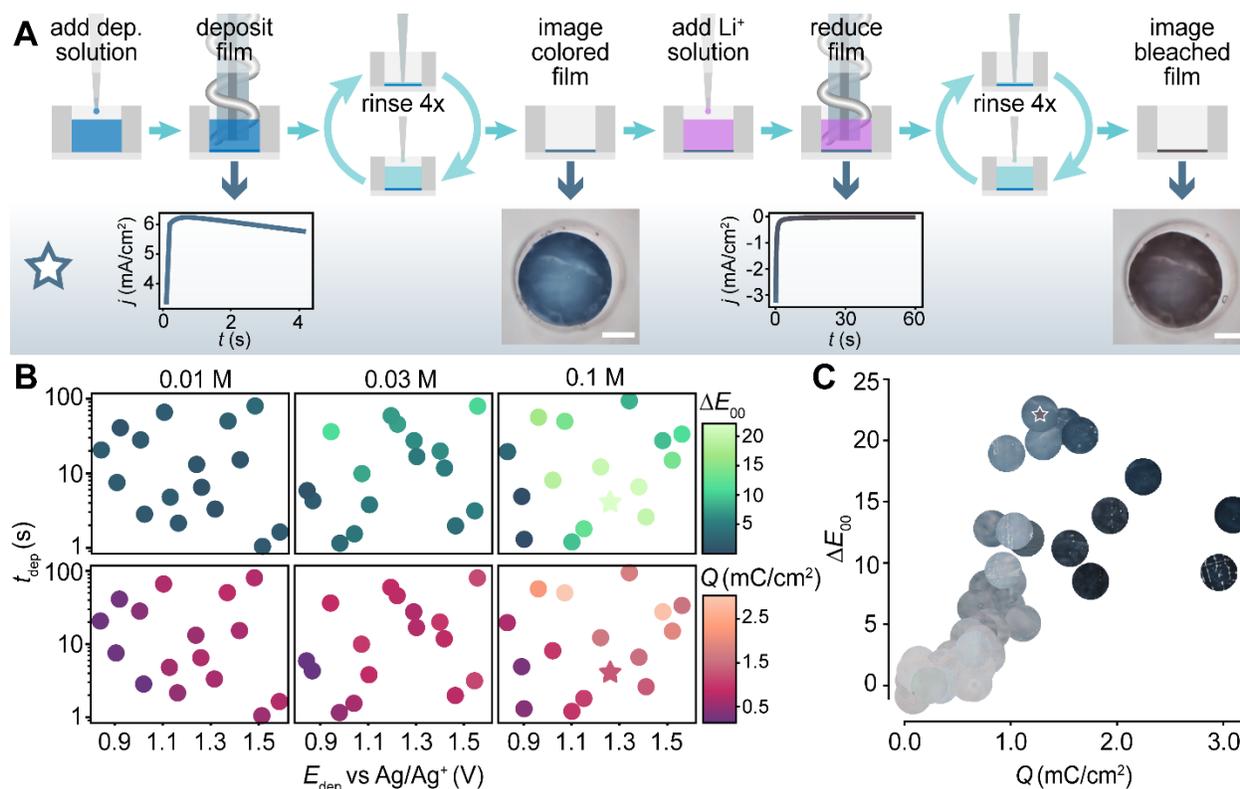

**Fig. 4** (A) Schematic showing the experimental sequence for depositing and characterizing of a poly(3,4-ethylenedioxythiophene) (PEDOT) film. Dark blue arrows indicate the workflow. Scale bars are 2 mm. (B) Plots comparing deposition parameters (potential $E_{dep}$ vs. Ag/Ag⁺ and deposition time $t_{dep}$) for 3,4-ethylenedioxythiophene (EDOT) solutions with concentrations of 0.01, 0.03, and 0.1 M. Marker color represents the colored/bleached contrast $\Delta E_{00}$ and total charge passed $Q$ during reduction. (C) Plot of $\Delta E_{00}$ vs. $Q$ for all data shown in B. Images at each point show optical images of the colored films. The points denoted by stars in B and C are the best-performing experiment and this data is also shown in A.

In considering how to explore the processing space associated with PEDOT:PSS, we identified three variables that would be pertinent to the final structure. Initially, we hypothesized that electrodeposition time, electrodeposition voltage, and EDOT concentration would be the most impactful three variables. Thus, to establish a training dataset in this three-dimensional space, we employed Latin hypercube sampling (LHS) to select 16 distinct experiments at each of three EDOT concentrations (0.01, 0.03, and 0.1 M), totaling 48 experiments. The parameters ranged from 0.8 to 1.6 V for the deposition potential and from 1 to 100 seconds for the deposition time, with the latter being selected in log space. We evaluated the outcomes by measuring the $\Delta E_{00}$, a color difference metric by CIE, reflecting the human-perceived color change (Figure 4B top), and by quantifying the total charge passed during film reduction with a $LiClO_4$ solution at -0.6 V for 60 s (Figure 4B bottom). We observed a trend where darker (thicker) films passed more charge during reduction, although films with moderate coloration exhibited the most significant color change between their colored and bleached states (Figure 4C).

**Autonomous experimentation with machine learning model**

While the prior experimental campaign showed that the PANDA was capable of autonomously depositing and functionally characterizing polymer films, these experiments were selected before starting the campaign, meaning that the experimental loop was open. Instead, the true value of SDLs is realized when each new experiment is selected based on the outcome of all prior experiments. Thus, we sought to show that the PANDA could be transformed into a true SDL by using machine learning to select each additional experiment. To test this concept, we implemented Bayesian optimization (BO) in which the data is modeled using a Gaussian process regression (GPR).

Leveraging the data from our prior campaign, we used the 48 LHS-selected data points as a training set to initially train the hyperparameters of the GPR using leave-one-out cross-validation. Noting that the optimal voltage was near the boundary, before proceeding to these BO-selected experiments, we expanded the range for our deposition parameters to 0.6 – 1.8 V for potential and included the 18 discrete EDOT concentrations (between 0.01 – 0.1 M) that could be obtained by mixing three stock solutions (0.01, 0.03, and 0.1 M) in 20 μL increments. The deposition time bounds remained the same (1 – 100 s). For each concentration, our ML model used LHS to generate 50,000 possible experiments, selecting parameters using maximum likelihood probability and expected improvement to optimize for $\Delta E_{00}$. These chosen parameters were then used by the robotic system to deposit and characterize a PEDOT film.

The PANDA active learning campaign proceeded by iteratively selecting an experiment using the GPR, performing the selected experiment, and then integrating the new results into the model in real time (Figure 5A). Due to our use of BO, the system naturally balanced exploring uncertain regions in parameter space and exploiting known high-performance areas. After 20 rounds of experimentation, the updated model predictions indicated a refined understanding of the parameter space, showing a realistic predicted measurement uncertainty (noise hyperparameter = 3.8) and identifying a reasonably smooth high-performing region which we assume to be the global maximum (Figure 5B).

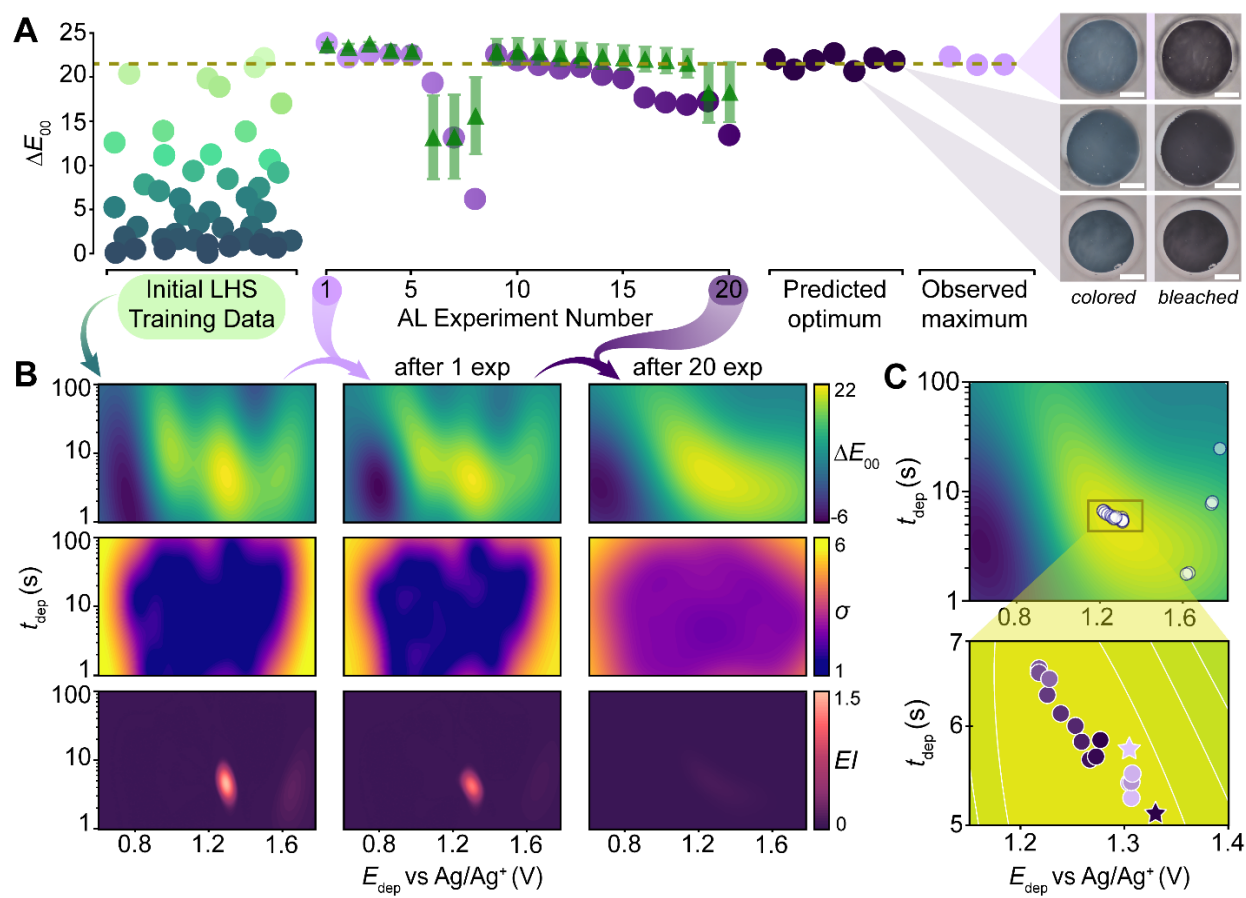

**Fig. 5** (A) Plot illustrating $\Delta E_{00}$ for initial LHS training data (left), autonomous learning (AL) data in purple (center), and validation points (right) shown in the order the experiments were performed. Model predictions and uncertainties are shown as green triangles with error bars and the dashed gold line indicates the model's predicted optimum after all experiments. Examples of champion films are shown in their colored and bleached states (right). (B) Surface plots of $t_{dep}$ vs. $E_{dep}$ showing (top) $\Delta E_{00}$ predicted by the GPR model, (middle) uncertainty $\sigma$ in the GPR model, and (bottom) computed expected improvement ($EI$). Data is shown after initial training data (left), after one AL experiment (middle), and after 20 AL experiments (right). (C) Magnified view of GPR-predicted $\Delta E_{00}$ after 20 experiments showing the AL data points. The bottom shows a magnified view of the maximum region with experiment order indicated by marker color. Stars indicate the parameters used for the validation experiments with the light purple showing the observed maximum and dark purple showing the predicted optimum.

In order to evaluate the effectiveness of the campaign, we performed a series of validation experiments. First, we queried the model to predict the parameter values expected to result in the highest $\Delta E_{00}$. These conditions were replicated in seven validation experiments shown in dark purple (Figure 5A predicted optimum), with the films being visually documented. Next, we retested the parameters from our observed maximum experiment during the active learning campaign by performing three additional replicates shown in light purple (Figure 5A observed maximum). While the validation experiments were performed outside the autonomous loop, the same evaluation code was used to determine their respective $\Delta E_{00}$ values. When running an active learning campaign, it is prudent to explore both the predicted and observed maximum as both can be suboptimal for different reasons. The predicted optimum can be incorrect if the model is not accurate in this region. The observed maximum can be suboptimal if the champion observed during the campaign happened to experience a large fluctuation. Here, the means of both overlapped with the predicted optimum showing $\Delta E_{00} = 21.7 \pm 0.7$ while the observed maximum resulted in $\Delta E_{00} = 22.2 \pm 1.2$. The agreement between these values is likely a representation of both the smoothness of this space together with the high quality of the model.

The experiments chosen by the model, overlayed on the model's refined parameter space, highlighted exploratory areas (higher voltages) and exploitative zones (around 1.2 – 1.3 V). A zoomed-in view of the area sampled the most by the model, between $5\text{ s} < t_{dep} < 7\text{ s}$, $1.15\text{ V} < E_{dep} < 1.4\text{ V}$, and 0.1 M EDOT concentration, shows the order experiments were performed in going from light (experiments performed first) to dark purple (experiments performed last) (Figure 5C). This illustrates the model testing the bounds of this smaller space before settling on a region that balanced both parameters.

## Conclusions

Taken together, we have demonstrated the use of a novel SDL for the synthesis and functional characterization of electrodeposited polymer films. Crucially, as this system was based on a new architecture, we performed extensive validation and calibration of the fluid handling and electrochemical subsystems. Finally, the autonomous optimization of PEDOT:PSS deposition conditions illustrates how this system can dynamically refine electrodeposition parameters. Given the modular and low-cost nature of this system, it has potentially broad applicability. Looking ahead, the full potential for SDLs in materials science is vast. As these systems become more accessible, bespoke SDL construction in a broader range of labs, including those with more restricted resources, can provide ultimately low-cost and efficient means of studying new materials systems. Ultimately, our findings support the continued development and investment in autonomous research platforms. As these systems evolve to become more sophisticated and user-friendly, they show promise of opening up new avenues in materials science and engineering, potentially altering how we approach and address some of the most pressing technological challenges.

## Author Contributions

Harley Quinn: data curation, formal analysis, investigation, methodology, resources, software, writing – original draft, validation, and visualization.

Gregory A. Robben: data curation, investigation, methodology, software, and writing – review and editing.

Zhaoyi Zhang: investigation, resources, writing – review and editing, and validation.

Alan L. Gardner: conceptualization, software, writing – review and editing.

Jörg G. Werner: conceptualization, funding acquisition, supervision, and writing – review and editing.

Keith A. Brown: conceptualization, funding acquisition, project administration, supervision, and writing – original draft.

## Conflicts of interest

There are no conflicts to declare.

## Acknowledgements

This work was supported by the National Science Foundation (CBET-2146597), the Toyota Research Institute, and the Boston University College of Engineering for support through the Dean's Catalyst Award. The authors are grateful to the Boston University Photonics Center and Engineering Product Innovation Center for providing access to instrumentation and resources critical to this work.

**PANDA: A self-driving lab for studying electrodeposited polymer films**

Harley Quinn,[a] Gregory A. Robben,[a] Zhaoyi Zheng,[a] Alan L. Gardner,[b] Jörg G. Werner,*[abc] and Keith A. Brown*[abd]

[a]Division of Materials Science & Engineering, Boston University, Boston, MA, 02215, USA.

[b]Department of Mechanical Engineering, Boston University, Boston, MA, 02215, USA

[c]Department of Chemistry, Boston University, Boston, MA, 02215, USA

[d]Department of Physics, Boston University, Boston, MA, 02215, USA

*E-mail: brownka@bu.edu, jgwerner@bu.edu

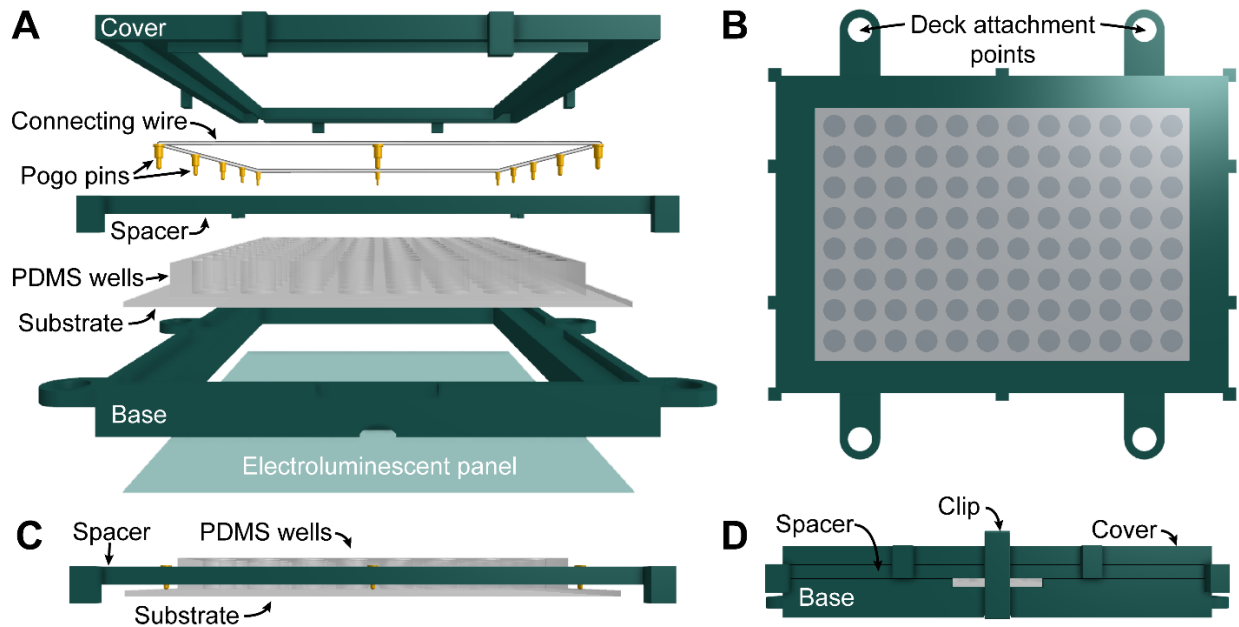

**Fig. S1** (A) Exploded view of substrate holder including the electroluminescent panel (27260-P, TechnoLight) used for optical transmission measurements. (B) Top view of holder with polydimethylsiloxane (PDMS) wells on substrate illustrating screw holes where it attaches to the deck. (C) Front view of substrate with PDMS wells showing the spacer and how gold-coated pogo (spring-loaded) pins contact the conductive surface of the transparent substrate. (D) Front view of fully assembled substrate holder with clip holding pieces together.

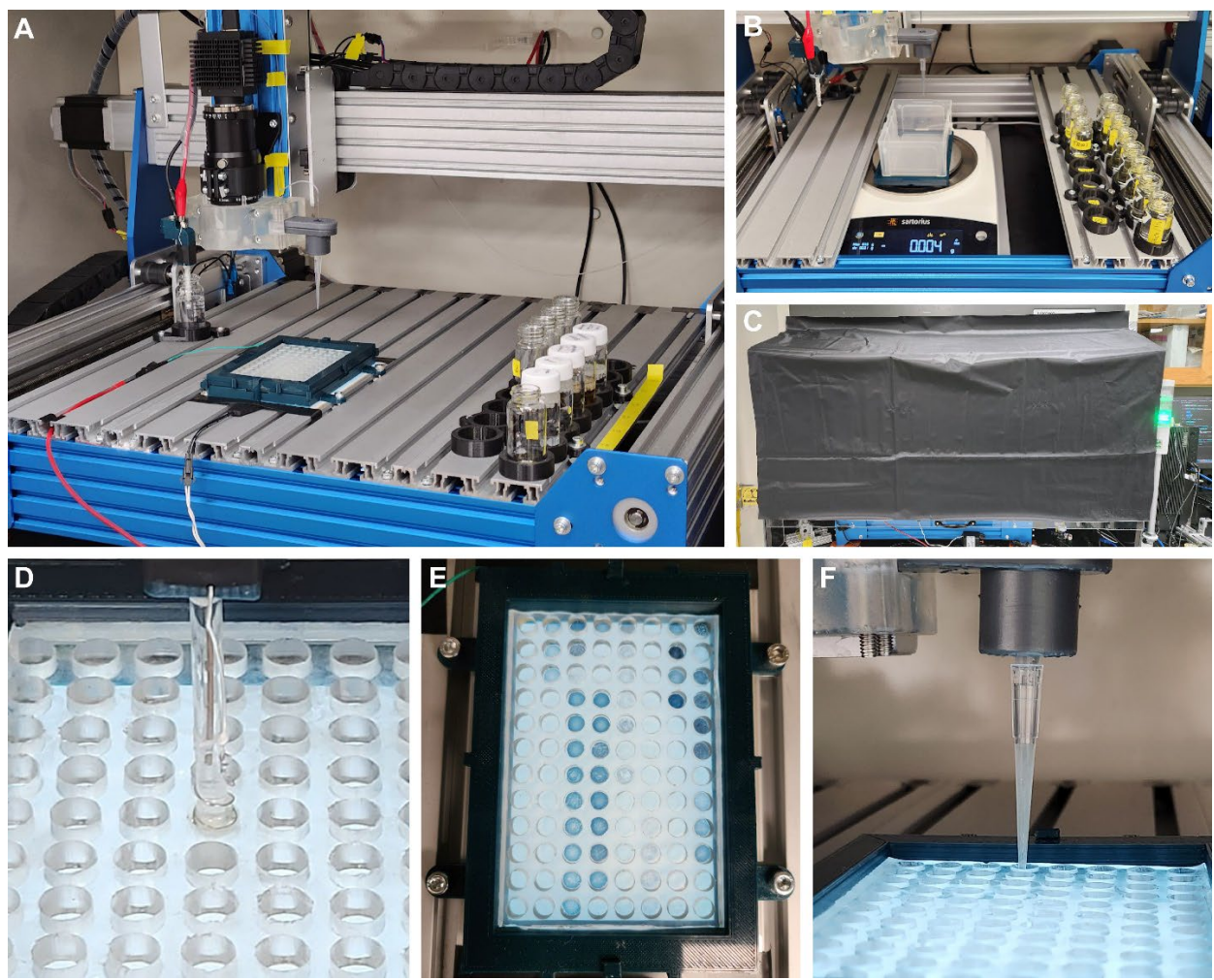

**Fig. S2** (A) Picture of the PANDA instrument deck with the reference electrode in electrolyte solution. (B) View of the PANDA deck with a panel removed for the scale, used in the liquid handling validation experiments. (C) PANDA enclosure with light blocking curtain installed for electrochromic experiments. (D) Close up view of counter and reference electrode in one of the PDMS wells. (E) Top view of the PDMS well plate over the electroluminescent panel with PEDOT:PSS films electrodeposited in some of the wells. (F) Close up view of the liquid handling system's pipette tip, removing solution from one of the wells after an experiment.

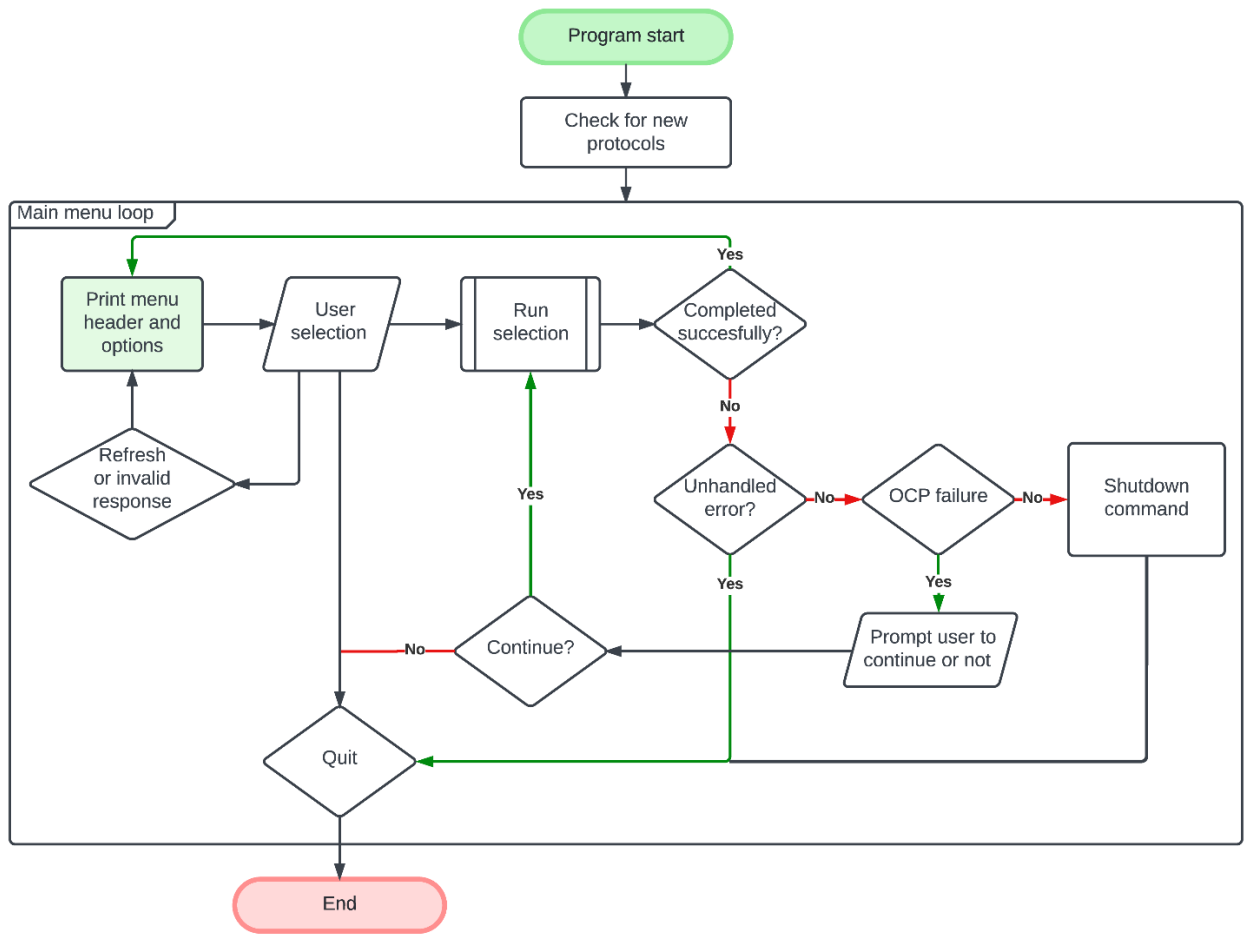

**Fig. S3** Flowchart of PANDA software main menu.

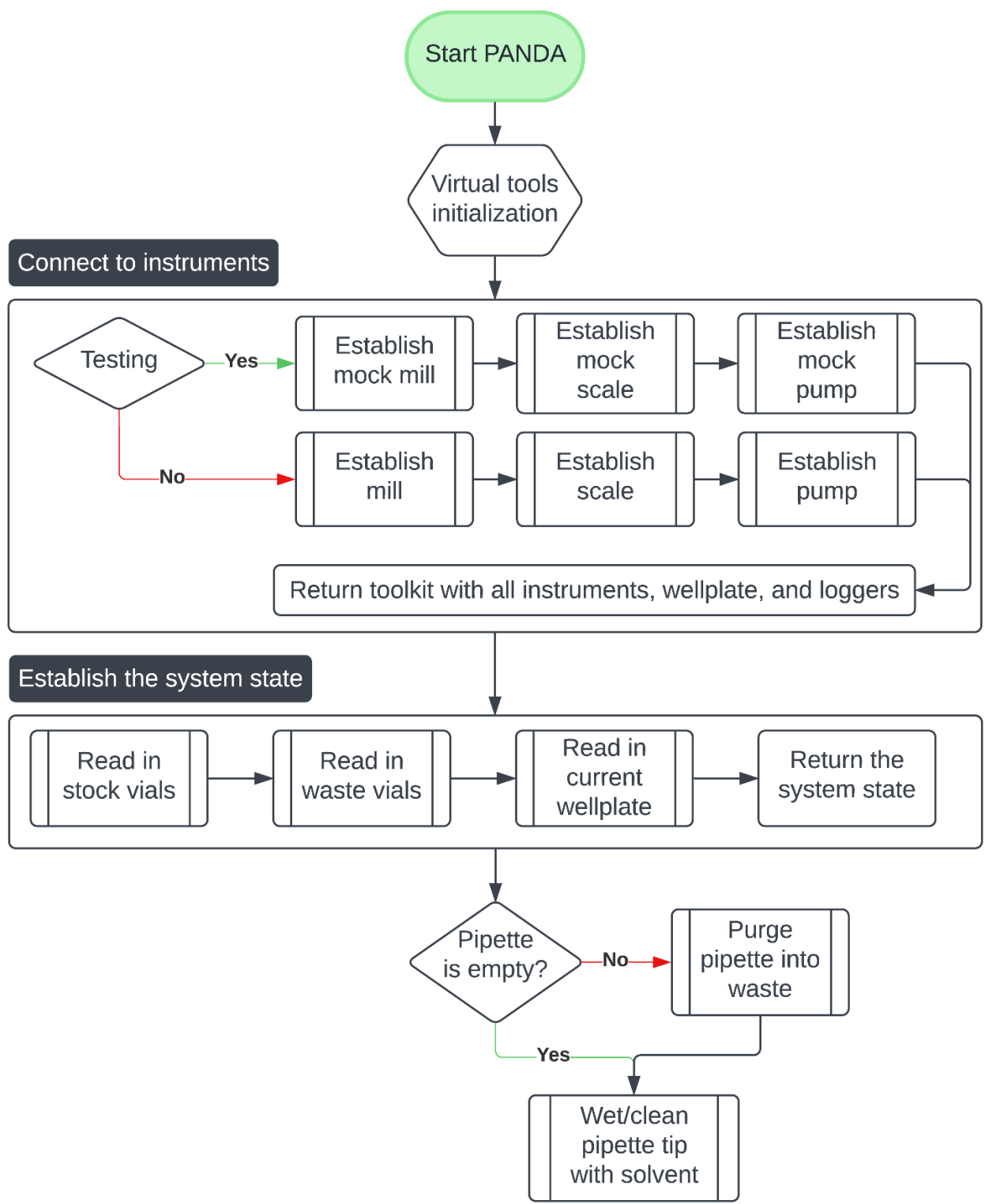

**Fig. S4** Flow chart showing initialization of PANDA prior to running the experiment loop.

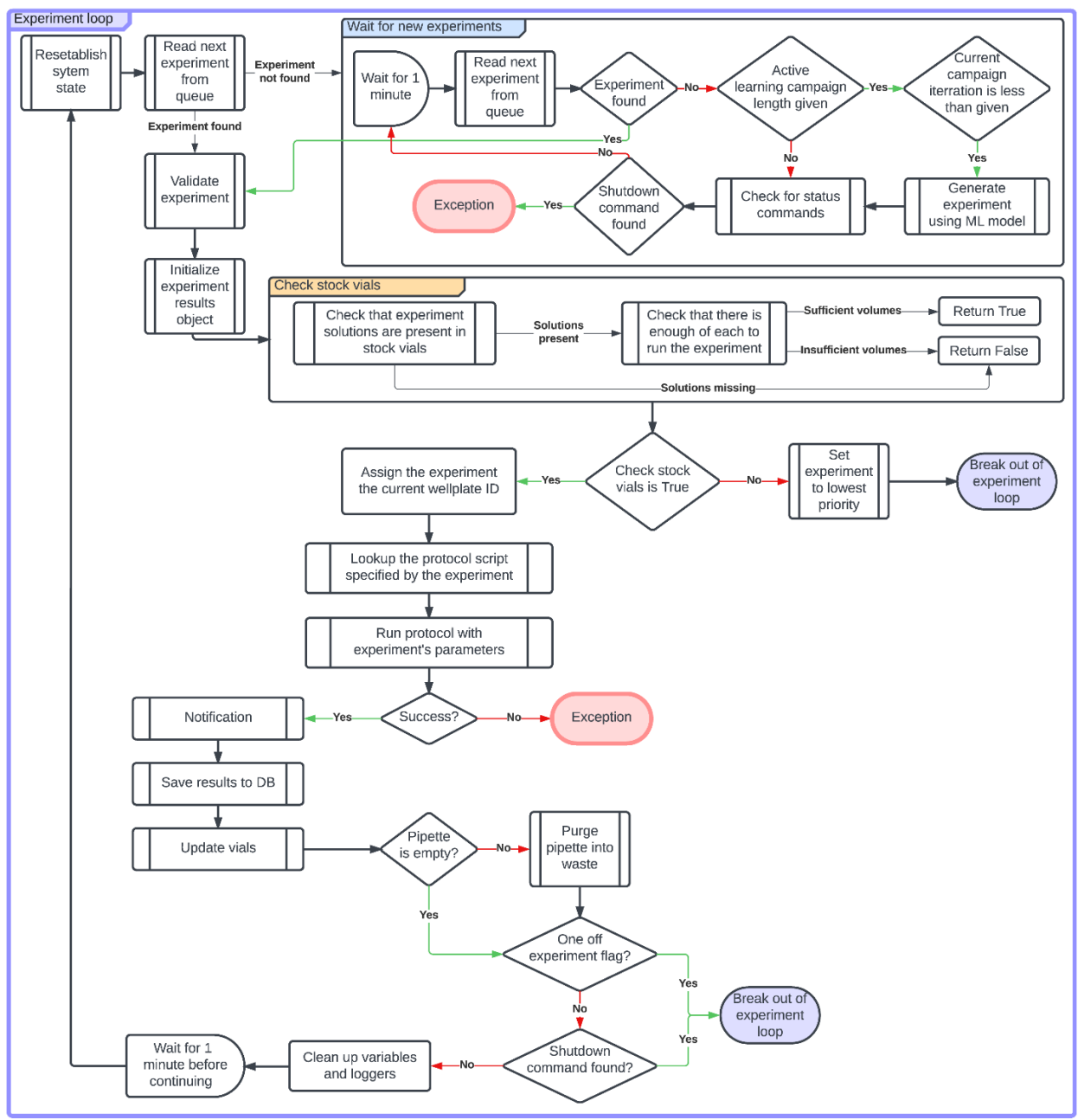

**Fig. S5** Flowchart of PANDA software showing experiment loop.

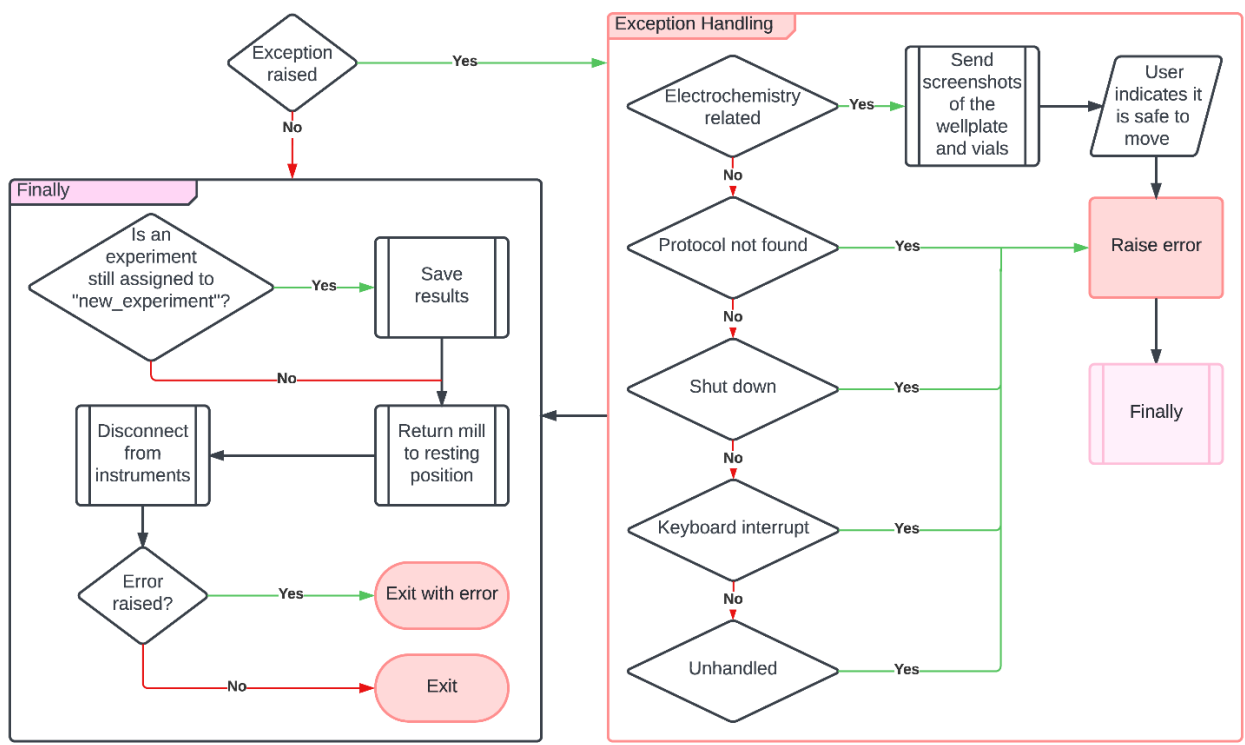

**Fig. S6** Flowchart of PANDA software demonstrating the end of the experiment loop, with error handling and shutting down the PANDA.

**Table S1** Open-source Python (version 3.11.9) packages used by the PANDA

| Purpose | Package Name | Version | Source |
|---|---|---|---|
| **Cameras** | obsws-python | 1.7.0 | https://pypi.org/project/obsws-python/ |
| **Machine Learning** | gpytorch | 1.11 | https://pypi.org/project/gpytorch/ |
| **Data visualization** | matplotlib | 3.8.4 | https://pypi.org/project/matplotlib/ |
| **Data handling** | pandas | 2.2.2 | conda-defaults |
| **Liquid handling calculations** | pulp | 2.8.0 | conda-forge |
| **Data validation** | pydantic | 2.5.3 | conda-defaults |
| **Instrument communication** | pyserial | 3.5 | conda-defaults |
| **Slack communications** | regex | 2023.10.3 | conda-defaults |
| **Image processing** | scikit-image | 0.23.2 | conda-defaults |
| **Machine learning** | scikit-learn | 1.4.2 | conda-defaults |
| **Slack communications** | slack-sdk | 3.19.5 | conda-defaults |

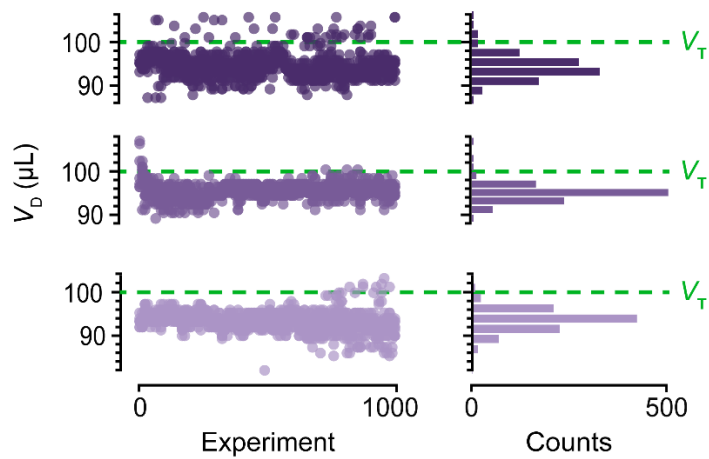

**Fig. S7** Plot showing three different pipette tips each performing 1000 fluid deposition experiments using forward pipetting, plotting experiment volume dispensed $V_D$ vs. experiment number. Histograms show the frequency of dispensed volumes. The target volume $V_T$ of 100 µL is marked by a dashed green line. RMSE for each pipette tip from top to bottom, 2.8 µL, 1.8 µL, and 2.2 µL.

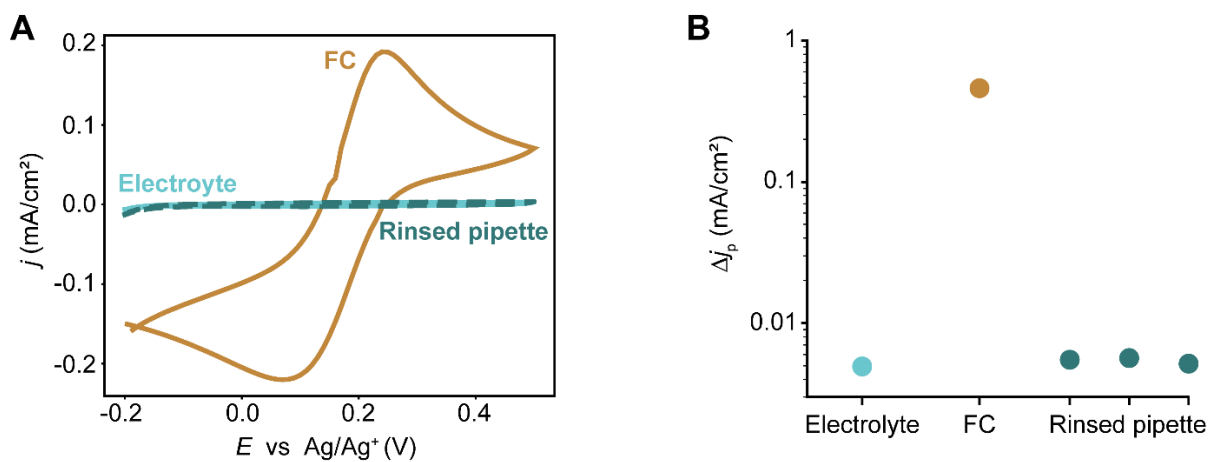

**Fig. S8** (A) Cyclic voltammograms (CVs) of an electrolyte solution (light teal), ferricyanide solution (FC), and electrolyte solution dispensed with a rinsed pipette (dark teal) performed with operating voltage between -0.2 and 0.5 V, scan rate 50 mV s$^{-1}$. (B) Plot showing the peak-to-peak current density $\Delta j_p$ values obtained for each CV from A.

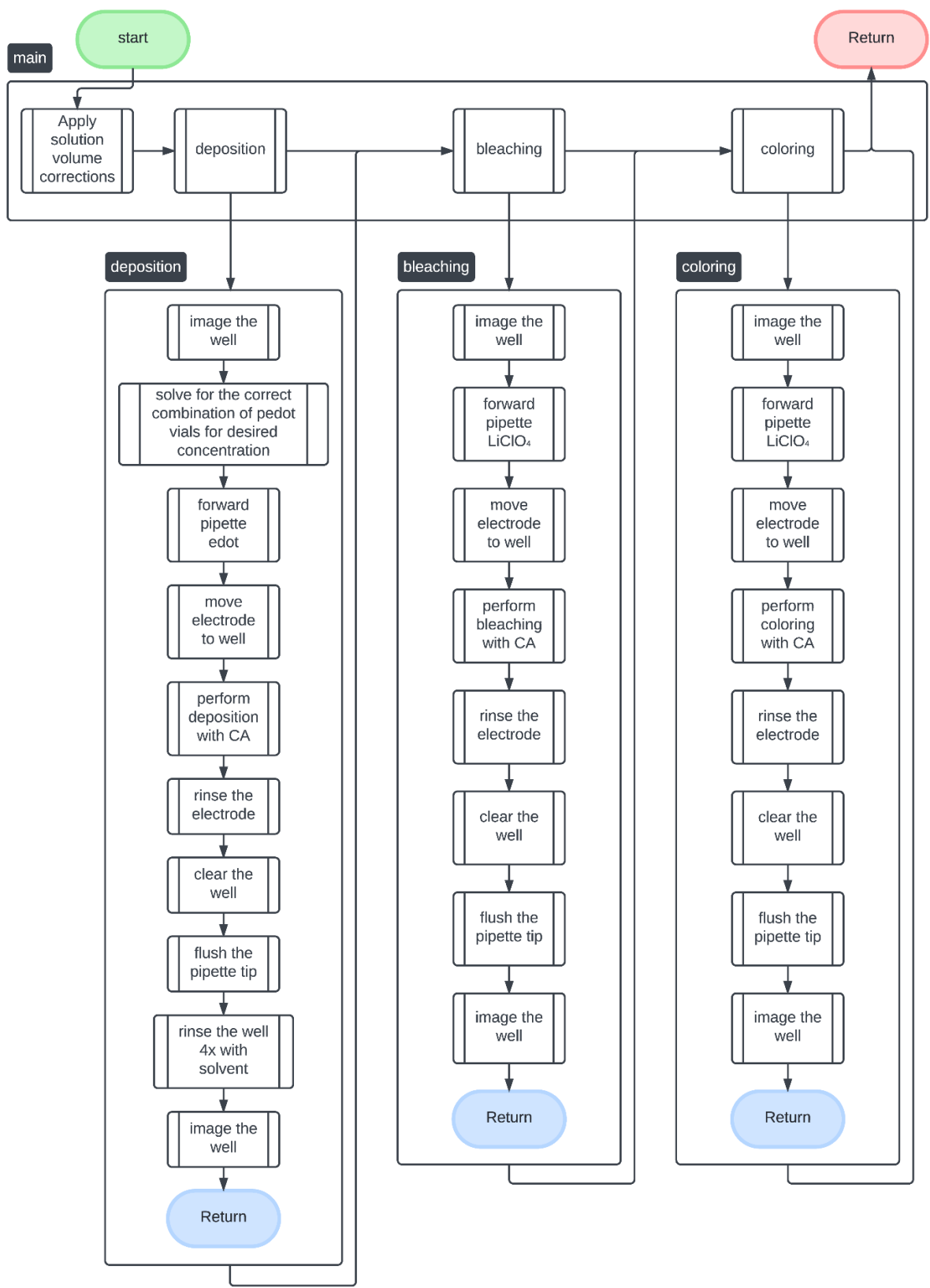

**Fig. S9** Flowchart illustrating the protocol for PEDOT:PSS experiments.